\documentclass[10pt,aps,preprintnumbers,nofootinbib,superscriptaddress,notitlepage,longbibliography]{revtex4-1}

\usepackage[T1]{fontenc}
\usepackage[latin1]{inputenc}
\usepackage[english]{babel}
\usepackage{stix}
\usepackage{epsfig}
\usepackage{amsmath,bm,bbm}
\usepackage{graphicx}
\usepackage{verbatim}
\usepackage{csquotes}
\usepackage{simplewick}
\usepackage{appendix}
\usepackage[usenames,dvipsnames,svgnames]{xcolor}
\usepackage[normalem]{ulem}
\usepackage{cleveref}

\newcommand{\eg}{\textit{e.g}}
\newcommand{\ie}{\textit{i.e}}
\newcommand{\dd}{\mathrm{d}}
\newcommand{\Mpl}{M_{\rm Pl}}

\begin{document}
\preprint{YITP-22-160, IPMU22-0069}
\title{Bouncing Cosmology in VCDM}
\author{Alexander Ganz}
\email{alexander.ganz@uj.edu.pl}
\affiliation{Faculty of Physics, Astronomy and Applied Computer Science, Jagiellonian University, 30-348 Krakow, Poland}
\author{Paul Martens}
\email{paul.martens@yukawa.kyoto-u.ac.jp}
\affiliation{Center for Gravitational Physics and Quantum Information (CGPQI),
Yukawa Institute for Theoretical Physics (YITP),
Kyoto University, 606-8502, Kyoto, Japan}
\author{Shinji Mukohyama}
\email{shinji.mukohyama@yukawa.kyoto-u.ac.jp}
\affiliation{Center for Gravitational Physics and Quantum Information (CGPQI),
Yukawa Institute for Theoretical Physics (YITP),
Kyoto University, 606-8502, Kyoto, Japan}
\affiliation{Kavli Institute for the Physics and Mathematics of the Universe (WPI),
The University of Tokyo, Kashiwa, Chiba 277-8583, Japan}
\author{Ryo Namba}
\email{ryo.namba@riken.jp}
\affiliation{RIKEN Interdisciplinary Theoretical and Mathematical Sciences (iTHEMS), Wako, Saitama 351-0198, Japan}
\date{\today}

\begin{abstract}
We construct an asymmetric bouncing scenario within the VCDM model\,---also known as type-II minimally modified gravity---\,, a modified gravity theory with two local physical degrees of freedom. The scenario is exempt of any ghost or gradient instability, 
ad-hoc matching conditions or anisotropic stress issue (BKL instability). It moreover succeeds in generating the cosmological perturbations compatible with the observations. The scalar spectral index can be adapted by the choice of the equation of state of the matter sector and the form of the VCDM potential leading to an almost scale-invariant power spectrum. Satisfying the CMB bounds on the tensor-to-scalar ratio leads to a blue tensor spectrum.
\end{abstract}

\maketitle

\section{Introduction}
\label{sec:introduction}

Inflation \cite{Guth:1980zm,Sato:1980yn,Starobinsky:1980te} has proven to be a very successful framework to simultaneously answer several major cosmological questions, \eg. the horizon problem, the flatness problem and the origin of primordial fluctuations. Its paradigm is robust enough to pass high-precision observational tests such as the one presented by the cosmological microwave background (CMB) \cite{Planck:2018vyg}. However, while phenomenologically satisfying, inflation also leaves us with a set of unanswered questions like the initial singularity \cite{Borde:1996pt,Borde:1993xh} and the trans-Planckian problem \cite{Brandenberger:2012uj,Brandenberger:2016vhg}. 

A popular alternative approach is the bouncing universe. That is a scenario of the universe where the cosmic expansion we are now observing was preceded by a contracting phase. The turning point between the two dynamics being called the \enquote{bounce}. In this case, the cosmic history is extended further in the past and gives a natural explanation for causal-connectedness. By introducing this pre-bounce history, the smoothness and flatness problems, as well as the horizon problem, are thus non-issues \cite{Brandenberger:2016vhg,Battefeld:2014uga,Cai:2014bea,Ijjas:2018qbo,Brandenberger:2012zb}. Therefore, a bouncing universe does not suffer from the aforementioned issues of inflation, while answering the same concerns the inflationary approach was built to address.

Noticeably, general relativity (GR) does not admit any bouncing solution under the null-energy condition. Therefore, if the Universe has to undergo a bounce, it must be described by an extended theory of gravity or by a non standard matter content. Several attempts have been made, within different frameworks to invoke such a cosmic history, in \eg. $f(T)$ gravity \cite{Cai:2011tc}, DHOST \cite{Zhu:2021whu,Ilyas:2020qja} or Ho\v{r}ava-Lifshitz gravity \cite{Brandenberger:2009yt}, using a quintom matter field \cite{Cai:2011zx}, a Cuscuton field \cite{Boruah:2018pvq,Kim:2020iwq}, and others. 

However, constructing viable bouncing models is a challenge. First, due to the violation of the null-energy condition these models tend to suffer commonly from ghost or gradient instabilities.
Within the Horndeski framework \cite{Deffayet:2011gz,Horndeski:1974wa,Kobayashi:2011nu}, that has led to a no-go theorem \cite{Kobayashi:2016xpl,Libanov:2016kfc}, and a similar result \cite{Vikman:2004dc} holds in $k$-essence models \cite{Armendariz-Picon:2000nqq,Armendariz-Picon:2000ulo} as well.
Nevertheless, these limitations have not prevented the development of a healthy bounce without ghost or gradient instability near the bounce \cite{Easson:2011zy,Cai:2012va}~\footnote{Another model based on the cubic Galileon action was also put forward \cite{Ijjas:2016tpn} with limitations \cite{Dobre:2017pnt} however.}. 

Alternatively, these issues can be avoided by working within more general frameworks like ghost condensation \cite{Creminelli:2006xe,Lin:2010pf} and beyond Horndeski/DHOST models \cite{Cai:2017dyi,Kolevatov:2017voe}, as also suggested by the effective field theory of cosmological perturbations \cite{Cai:2016thi,Cai:2017tku}. Another issue is the anisotropic stress, or the Belinski-Khalatnikov-Lifshitz (BKL) instability \cite{BELINSKII1992609}. Besides the conceptual problems the current observations set strict constraints on the scalar spectral index, $n_s \approx 0.96$, while the tensor-to-scalar ratio must respect the bounds of $r_{0.05} < 0.036$ ($95 \%$ CL) \cite{BICEP:2021xfz}. Naturally, a healthy bouncing scenario must account for these observations. However, while for instance the matter bounce is successful in obtaining an almost scale invariant power spectrum \cite{Wands:1998yp,Finelli:2001sr}, it breaks the bounds on the tensor-to-scalar ratio. Indeed, a conjectured no-go theorem \cite{Quintin:2015rta,Li:2016xjb,Akama:2019qeh} forbids a naive single scalar-field ($k$-essence CITE) matter bounce to simultaneously satisfy the requirement of a nearly scale-invariant scalar power spectrum, and the tensor-to-scalar ratio bounds, without producing excessive non-Gaussianities. Introducing additional scalar fields can reconcile the matter bounce via the curvaton mechanism \cite{Cai:2011zx}.

In the present study, we exhibit a full and concrete model of a bouncing universe scenario, built within the formalism of the minimally modified gravity (MMG) \cite{Lin:2017oow,Mukohyama:2019unx,DeFelice:2020eju}. These theories do not introduce additional local physical degrees of freedom other than those in GR, while they may contain global modes called shadowy modes (or generalized instantaneous modes)~\footnote{See \cite{DeFelice:2018ewo,DeFelice:2021hps} for shadowy modes in the context of U-DHOST theories.} due to the existence of a preferred frame. Therefore, they easily avoid instabilities and constraints that could stem from extra propagating degrees of freedom that are common in other modified gravity theories, even without needing any screening mechanisms. In our work, we will consider the VCDM model, a specific type of MMG theory \cite{DeFelice:2020eju}. According to the classification introduced in \cite{Aoki:2018brq}, the VCDM is a type-II MMG theory since it has no Einstein frame~\cite{Aoki:2021zuy}.
The name \enquote{VCDM} comes from promoting the cosmological constant $\Lambda$ of the standard $\Lambda$CDM model to a function $V(\varphi)$ of a non-dynamical, auxiliary field $\varphi$.
Extending its original usage for the late-time universe, various aspects of the VCDM, including attempts to address tensions in late-time cosmology \cite{DeFelice:2020cpt,DeFelice:2020prd}, black holes \cite{DeFelice:2020onz}, stars \cite{DeFelice:2021xps}, gravitational collapse \cite{DeFelice:2022riv} and the solution space including GR solutions \cite{DeFelice:2022uxv}, have been explored. When applied to the very early universe, the VCDM model has the advantage, by construction, to provide the freedom to realize this bouncing scenario as well as safely return to GR after the bounce. It evades the aforementioned no-go theorems, yet provides just enough of a framework to violate the null energy condition, similarly to what was shown recently with Cuscuton \cite{Boruah:2018pvq}. As recently shown in \cite{DeFelice:2022uxv}, any solutions of the Cuscuton model \cite{Afshordi:2006ad} are solutions of the VCDM model. However, as shown in the same paper \cite{DeFelice:2022uxv}, the VCDM also admits other solutions, such as GR solutions. Furthermore, the framework of VCDM greatly simplifies the reconstruction of the potential in the Lagrangian from background cosmological histories, as already shown in \cite{DeFelice:2020eju} for general expanding backgrounds. Within the present study, we shall consider whether the scalar power spectrum is (almost) scale invariant at superhorizon scales, as well as investigate the tensor-to-scalar ratio, so that these observables are indeed compatible with the observations.

Our paper is organized as follows. In \cref{sec:vcdm}, the VCDM model is introduced under the ADM decomposition. The general formulations of its background and linear perturbations of the tensor, vector and scalar modes are derived in \cref{sec:background,sec:linear}, respectively. In \cref{sec:bouncing}, a concrete bouncing dynamics is implemented in the model, and the predictions of the scalar and tensor power spectra are computed. \Cref{sec:conclusion} is devoted to discussions and conclusions of the paper.

\section{VCDM}
\label{sec:vcdm}

The construction of this class of theory is based on the ADM decomposition of the $4$-D metric into the time slice and the spatial hypersurface as
\begin{equation}
  \label{eq:4}
  \dd s^2 = - N^2 \dd t^2 + \gamma_{ij} \left( N^i \dd t + \dd x^i \right) \left( N^j \dd t + \dd x^j \right) \; ,
\end{equation}
where $N$ and $N^i$ are respectively the lapse function and shift vector, and the metric $\gamma_{ij}$ describes the $3$-D spatial manifold.
We can then define a vector $n^\mu$ normal to time-constant hypersurfaces such that
\begin{equation}
  \label{eq:5}
  n^\mu \equiv \left( \frac{1}{N} , \, - \frac{N^i}{N} \right) \; ,
\end{equation}
and the (inverse) spatial metric extended to $4$-D such that
\begin{equation}
  \label{eq:6}
  h^{\mu\nu} \equiv
  \left(
  \begin{array}{cc}
    0 & 0 \\
    0 & \gamma^{ij}
  \end{array}
  \right) \; .
\end{equation}
Note that the $4$-D metric can be decomposed by
\begin{equation}
  \label{eq:24}
  g^{\mu\nu} = h^{\mu\nu} - n^\mu n^\nu \; .
\end{equation}
The temporal derivative of $\gamma_{ij}$ appears in the combination of the extrinsic curvature
\begin{equation}
  \label{eq:7}
  K_{\mu\nu} \equiv \frac{1}{2} \, \pounds_n h_{\mu\nu}
  = \frac{1}{2} \left( n^\rho \partial_\rho h_{\mu\nu} + h_{\mu\rho} \partial_\nu n^\rho + h_{\rho\nu} \partial_\mu n^\rho \right)
  = \frac{1}{2} \left( n^\rho \nabla_\rho h_{\mu\nu} - n^\rho \nabla_\nu h_{\mu\rho} - n^\rho \nabla_\mu h_{\rho\nu} \right) \; ,
\end{equation}
and its spatial projection reads
\begin{equation}
  \label{eq:8}
  K_{ij} = \frac{1}{2N} \left( \partial_t \gamma_{ij} - D_i N_j - D_j N_i \right) \; ,
\end{equation}
where $\pounds_n$ and $D_i$ is the Lie derivative along $n^\mu$ and covariant derivative associated with $\gamma_{ij}$ and $N_i \equiv \gamma_{ij} N^j$, respectively.

Using these definitions, the action for VCDM can be written as
\begin{equation}
  \label{action_mmg}
  S_\mathrm{VCDM} = \frac{\Mpl^2}{2} \int \dd^4x N \sqrt{\gamma}
  \left[
  \mathcal{R} + K_{ij} K^{ij} - K^2 - 2 V(\varphi)
  - 2 \, \frac{\lambda^i}{N} \, \partial_i \varphi - \frac{3}{2} \, \lambda_0^2 - 2 \lambda_0 \left( K + \varphi \right)
  \right] \; ,
\end{equation}
where $K \equiv \gamma^{ij} K_{ij}$, $\mathcal{R}$ is the Ricci scalar associated with $\gamma_{ij}$, the quantities $\lambda^i$ and $\lambda_0$ are Lagrange multipliers, and $\varphi$ is an auxiliary scalar field. Because of its non-trivial constraint structure, this theory contains only $2$ propagating degrees of freedom (dof); that is the same number as GR. Now, we introduce a matter field that evolves on the background. This field is here modelized by a (shift-symmetric) $k$-essence type of field, explicitly
\begin{align}
  \label{action_matter}
  S_\mathrm{matter} = \Mpl^2 \int \dd^4x N \sqrt{\gamma} \, P(X) \; , \qquad
  X \equiv - \frac{1}{2} \, g^{\mu\nu} \partial_\mu \chi \, \partial_\nu \chi
  = \frac{1}{2} \left[ \left( \partial_\perp \chi \right)^2 - \gamma^{ij} \partial_i \chi \, \partial_j \chi \right] \; ,
\end{align}
where $\partial_\perp \chi \equiv n^\mu \partial_\mu \chi$. We have chosen here to normalize the matter sector such that $\Mpl^2$ multiplies the entire matter action. The total action is thus
\begin{equation}
  \label{action_total}
  S = S_\mathrm{VCDM} + S_\mathrm{matter} \; .
\end{equation}
Since the VCDM alone has only 2 (tensor) dof's, the introduction of the matter sector is essential to generate scalar perturbations, which eventually seed the structure formation in the universe.

\section{Background bouncing solutions}
\label{sec:background}

To consider a homogeneous and isotropic background, we take the following background quantities
\begin{align}
  \begin{aligned}
    \label{BG_metric}
    N & = \bar{N}(t) \; ,
      & N^i & = 0 \; ,
      & \gamma_{ij} & = a^2(t) \, \delta_{ij} \; ,
      & \varphi & = \phi(t) \; , \\
     \qquad \lambda^i & = 0 \; ,
      & \lambda_0 & = \bar\lambda(t) \; ,
      & \chi & = \bar\chi(t) \; .
  \end{aligned}
\end{align}
Then the variations of the action \eqref{action_total} with respect to $\bar N, \, \bar\lambda, \, \phi, \, a, \, \bar\chi$ lead, respectively, to
\begin{subequations}
  \begin{align}
    \label{eq:3}
    3 H^2 & = V + \phi \bar \lambda + \frac{3}{4} \, \bar \lambda^2 + 2 X P_X - P \; , \\
    0 & = 3 H + \phi + \frac{3}{2} \, \bar\lambda \; , \\
    0 & = \bar\lambda + V_\varphi \; , \\
    2 \, \frac{\partial_t H}{\bar N} + 3 H^2 & = V + \phi \bar\lambda + \frac{3}{4} \, \bar\lambda^2 - \frac{\partial_t \bar\lambda}{\bar N} - P \; , \\
    0 & = \left( P_X + 2 X P_{XX} \right) \frac{1}{\bar N} \, \partial_t \left(  \frac{\partial_t \bar\chi}{\bar N} \right) + 3H P_X \, \frac{\partial_t \bar\chi}{\bar N} \; ,
  \end{align}
\end{subequations}
where $H \equiv \partial_t a / (aN)$, and $V, X$ and $P$ (and their derivatives) are all evaluated at the background values. By manipulating the above equations, they can be rewritten in a more convenient form as
\begin{subequations}
  \begin{align}
    \label{eq:const_1}
    0 & = V - \frac{\phi^2}{3} + \rho_\chi \; , \qquad \rho_\chi \equiv 2 X P_X - P \; ,\\
    \label{eq:const_2}
    0 & = 3 H + \phi - \frac{3}{2} \, V_\varphi \; , \\
    \label{eq:const_3}
    0 & = \bar\lambda + V_\varphi \; , \\
    \label{eq:2ndFriedmann}
    2 \, \frac{\partial_t H}{\bar N} & = 
    V_{\varphi\varphi} \, \frac{\partial_t \phi}{\bar N} - \left( \rho_\chi + P \right) \; , \\
    \label{eq:eom_chi}
    0 & = \frac{1}{\bar N} \, \partial_t \left(  \frac{\partial_t \bar\chi}{\bar N} \right) + 3 c_s^2 H \, \frac{\partial_t \bar\chi}{\bar N} \; , \qquad
    c_s^2 \equiv \frac{P_X}{P_X + 2 X P_{XX}} \; ,
  \end{align}
\end{subequations}
provided that $P_X + 2 X P_{XX} \ne 0$.

Combining the time derivative of \cref{eq:const_2,eq:2ndFriedmann} in the above expressions, we find
\begin{equation}
  \label{eq:10}
  \frac{\partial_t \phi}{\bar N} = 3 P_X X = \frac{3}{2} \left( \rho_\chi + P \right) \; .
\end{equation}
Also note that, as standard, \cref{eq:eom_chi} can be rewritten as
\begin{equation}
  \label{eq:9}
  \frac{\partial_t \rho_\chi}{\bar N} + 3 H \left( \rho_\chi + P \right) = 0 \; .
\end{equation}
Combining these last two expressions, \cref{eq:10,eq:9}, one can formally write
\begin{equation}
  \label{phi_formalsol}
  \phi = \frac{3}{2} \int^t \bar{N} \dd t' \left( \rho_\chi + P \right)
  = - \frac{1}{2} \int^t \dd t' \, \frac{\partial_{t'} \rho_\chi}{H}
  \;.
\end{equation}

For our purpose, we consider from now on a matter species with a constant equation of state $w \equiv P / \rho_\chi = {\rm const.}$. This can be realized by choosing $P(X)$ as
\begin{equation}
  \label{P_form_target}
  P = P_0 \, X^{\frac{1+w}{2w}} = P_0 \, X^{\frac{\gamma}{2(\gamma-3)}} \; , \qquad
  \gamma \equiv 3(1+w) \; ,
\end{equation}
where $P_0$ is some constant.
Then we observe the energy density of $\chi$ behaves as a matter with equation of state $w$, \ie.,
\begin{equation}
  \label{rho_target}
  \rho_\chi = \rho_0 \left( \frac{a_0}{a} \right)^\gamma
   \; ,
\end{equation}
where subscript $0$ denotes values at some fiducial time.

\section{Linear perturbations}
\label{sec:linear}

We now consider perturbations around the background \cref{BG_metric}. We expand the lapse, shift and $3$-D metric as
\begin{equation}
  \label{eq:11}
  N = \bar{N}(t) \left( 1 + \nu \right) \; , \qquad
  N^i = \frac{\bar{N}(t)}{a(t)} \left( \partial_i \beta + B_i \right) \; , \qquad
  \gamma_{ij} = a^2(t) \, {\rm e}^{2 \zeta} \left[ \delta_{ij} + 2 \partial_i \partial_j E + 2 \partial_{(i} E_{j)} + h_{ij} + \frac{1}{2} \, h_{ik} h_{kj}  \right] \; ,
\end{equation}
where $\{ \nu, \beta, \zeta, E \}$ are scalar perturbations, $\{ B_i, E_i \}$ are vectors ($\partial_i B_i = \partial_i E_i = 0$), and $\{ h_{ij} \}$ are tensors ($\partial_i h_{ij} = h_{[ij]} = h_{ii} = 0$), and they all depend on both time and space coordinates. We also expand the auxiliary fields $\{ \varphi , \lambda_0 , \lambda^i \}$ and the matter field $\chi$ as
\begin{equation}
  \label{eq:15}
  \varphi = \phi(t) + \delta\varphi(t, \bm{x}) \; , \qquad
  \lambda_0 = \bar\lambda(t) + \delta\lambda_0(t, \bm{x}) \; , \qquad
  \lambda^i = \frac{1}{a^2} \left[ \partial_i \delta\lambda_s(t, \bm{x}) + \delta\lambda_i(t, \bm{x}) \right] \; , \qquad
  \chi = \bar\chi(t) + \delta\chi(t, \bm{x}) \; ,
\end{equation}
where $\partial_i \delta\lambda_i = 0$. The theory \cref{action_mmg} under consideration does not respect the symmetry under the temporal coordinate transformation but still preserves the spatial diffeomorphism. Under the transformation
\begin{align}
  \label{eq:14}
  x^i \to x^i + \xi^i(\bm{x}) \; ,
\end{align}
each variable transforms by the amount, at the linear order,
\begin{equation}
  \begin{gathered}
    \label{transform_3d}
    \Delta E = a^2 \xi_L \; , \qquad
    \Delta E_i = a^2 \xi^i_T \; ,
    \\
    \Delta\nu = \Delta\beta = \Delta\zeta = \Delta B_i = \Delta h_{ij} = \Delta\delta\varphi = \Delta\delta\lambda_0 = \Delta \delta\lambda_s = \Delta\delta\lambda_i = \Delta\delta\chi = 0 \; ,
  \end{gathered}
\end{equation}
where $\xi^i$ has been expanded as
\begin{equation}
  \label{eq:20}
  \xi^i = \partial_i \xi_L + \xi^i_T \; , \qquad
  \partial_i \xi^i_T = 0 \; .
\end{equation}
As can be seen, the $h_{ij}$ components are gauge-invariant, as in GR. Additionally, $\nu$ and $\zeta$ are also independent of the $3$-D spatial gauge choice.%
\footnote{
  For $4$-D transformation $x^\mu \to x^\mu + \xi^\mu$, writing $\xi_0 \equiv \bar{N} \xi^0$, the variables transform as
  \begin{equation}
    \label{transform_4d}
    \Delta\nu = \frac{\partial_t \xi_0}{\bar{N}} \; , \quad
    \Delta\beta = \frac{a}{\bar{N}} \, \partial_t \xi_L - \xi_0 \; , \quad
    \Delta\zeta = H \xi_0 \; , \quad
    \Delta E = a^2 \xi_L \; , \quad
    \Delta B_i = \frac{a}{\bar{N}} \, \partial_t \xi^i_T \; , \quad
    \Delta E_i = a^2 \xi^i_T \; , \quad
    \Delta h_{ij} = 0 \; , \quad
    \Delta\delta\chi = \frac{\partial_t \bar\chi}{\bar{N}} \, \xi_0 \; .
  \end{equation}
}
We now use the freedom of $\xi_L$ and $\xi^i_T$ to fix the gauge by setting
\begin{align}
  \label{eq:21}
  E = E_i = 0 \; , \qquad
  \mbox{gauge choice} \; .
\end{align}
Then we work through the calculations for the following variables:
\begin{align*}
  & \mbox{Scalar modes: } \; \nu , \, \beta , \, \zeta , \, \delta\varphi , \, \delta\lambda_0 , \, \delta\lambda_s , \, \delta\chi \; , \\
  & \mbox{Vector modes: } \; B_i , \, \delta\lambda_i \; , \\
  & \mbox{Tensor modes: } \; h_{ij} \; ,
\end{align*}
among which $\{ \nu, \beta, \delta\varphi, \delta\lambda_0 , \delta\lambda_s \}$ and $\{ B_i , \delta\lambda_i \}$ are non-dynamical modes (\ie.~they appear in the action without time derivatives, up to total derivatives). On top of that, due to the peculiar constraint structure of MMG, one of the remaining scalar degrees of freedom is also non-dynamical. Therefore, at the end of the day, we have the following number of propagating (dynamical) degrees of freedom:
\begin{align*}
  \mbox{Scalar: } & \; 1 \; \mbox{dof} \; , \\
  \mbox{Vector: } & \; 0 \; \mbox{dof  (all non-dynamical)} \; , \\
  \mbox{Tensor: } & \; 2 \; \mbox{dof} \; .
\end{align*}
This counting is the same as in GR ($+$ one matter dof). Subsequently, we perform the perturbative analysis of the quadratic action for each sector separately.

\subsection{Tensor sector}
\label{subsec:tensor}

In the following, we use the conformal time $\tau$ (akin to setting $\bar N=a$). The tensor sector $\{ h_{ij} \}$ is essentially the same as GR. Decomposing $h_{ij}$ into polarization modes in the Fourier space, it reads
\begin{equation}
  \label{eq:25}
  h_{ij}(\tau,\bm{x}) = \sum_{\sigma} \int \frac{\dd^3 k}{(2\pi)^{3/2}} \, {\rm e}^{i \bm{k} \cdot \bm{x}} \, \Pi_{ij}^\sigma \big(\hat{k}\big) \, {h}_\sigma(\tau, \bm{k}) \; ,
\end{equation}
where we now used the conformal time $\tau$, and where $\Pi_{ij}^{\sigma}$ is the polarization tensor for the $2$ polarization modes, satisfying
\begin{equation}
  \label{eq:27}
  \delta^{ij} \Pi_{ij}^\sigma \big( \hat{k} \big) = \hat{k}^i \Pi_{ij}^\sigma \big( \hat{k} \big) = 0 \; , \qquad
  \Pi_{ij}^{\sigma} \big( \hat{k} \big) \, \Pi_{ij}^{\sigma' \, *} \big( \hat{k} \big) = \delta^{\sigma \sigma'} \; , \qquad
  \Pi_{ij}^{\sigma \, *}\big( \hat{k} \big) = \Pi_{ij}^\sigma \big( - \hat{k} \big) \; ,
\end{equation}
and these modes are decoupled at the linear order. Thanks to these properties and the reality condition of $h_{ij}(t, \bm{x})$, we see ${h}_\sigma^\dagger(\bm{k}) = {h}_{\sigma} (- \bm{k})$. Then the quadratic action for ${h}_\sigma(\tau,\bm{k})$ reads
\begin{equation}
  \label{eq:26}
  S^{(2)}_T = \frac{\Mpl^2}{8} \sum_\sigma \int  \dd \tau \, \dd^3k \, a^2
  \left[
   \vert {h}_\sigma^\prime \vert^2
  - k^2 \, \vert {h}_\sigma \vert^2
  \right] \; .
\end{equation}
where the prime ($^\prime$) denotes a derivative with respect to conformal time.
To obtain this, there is no use of background equations. The tensor sector is as standard as GR.

\subsection{Vector sector}
\label{subsec:vector}

The vector sector $\{B_i , \delta\lambda_i \}$ is as trivial as in GR. In fact $\delta\lambda_i$ simply does not appear in the quadratic action. We thus decompose $B_i$ into polarizations in the Fourier space,
\begin{equation}
  \label{eq:28}
  B_i(t,\bm{x}) = \sum_{s} \int \frac{\dd^3k}{(2\pi)^{3/2}} \, {\rm e}^{i \bm{k} \cdot \bm{x}} \, \epsilon_i^s \big( \hat{k} \big) \, {B}_s( \tau, \bm{k}) \; ,
\end{equation}
where $\epsilon_i^s$ is the polarization vector satisfying
\begin{equation}
  \label{eq:29}
  \hat{k}^i \epsilon_i^s\big( \hat{k} \big) = 0 \; , \qquad
  \epsilon_i^s \big( \hat{k} \big) \, \epsilon_i^{s' \, *} \big( \hat{k} \big) = \delta^{s s'} \; , \qquad
  \epsilon_i^{s \, *} \big( \hat{k} \big) = \epsilon_i^s \big( - \hat{k} \big) \; ,
\end{equation}
and the reality condition of $B_i(\tau ,\bm{x})$ results in ${B}_s^\dagger (\bm{k}) = {B}_s ( - \bm{k} )$. The quadratic action for the vector sector then reads
\begin{equation}
  \label{eq:30}
  S^{(2)}_V = \frac{\Mpl^2}{4} \int  \dd \tau \,  \dd^3 k \, a^2 k^2 \,  \vert {B}_s \vert^2 \; .
\end{equation}
Therefore there is no dynamical vector mode, just like in GR.

\subsection{Scalar sector}
\label{subsec:scalar}

The scalar sector $\{ \nu, \beta , \delta\varphi , \delta\lambda_0 , \delta\lambda_s , \zeta , \delta\chi \}$ is the non-trivial one. Let us first Fourier-decompose each variable as
\begin{equation}
  \label{eq:31}
  \delta(t,\bm{x}) = \int \frac{\dd^3k}{(2\pi)^{3/2}} \, {\rm e}^{i \bm{k} \cdot \bm{x}} \, \delta(\tau, \bm{k}) \; ,
\end{equation}
where $\delta = \{ \nu, \beta , \delta\varphi , \delta\lambda_0 , \delta\lambda_s , \zeta , \delta\chi \}$. Note the reality condition imposes $\delta^\dagger (\bm{k}) = \delta(- \bm{k})$. In order to eliminate the non-dynamical variables in favor of the dynamical ones, we employ the Faddeev-Jackiw method \cite{Faddeev:1988qp}. Due to the non-trivial structure of the theory, we need to impose the background equations before integrating out the non-dynamical variables, in order to obtain all the constraint equations. As counted at the beginning of this section, there is only $1$ dynamical degree of freedom. We have some freedom to choose the variable we wish to work with. It is convenient to choose the comoving curvature perturbation, defined as 
\begin{equation}
  \label{eq:def_R}
  \mathcal{R}_k \equiv \zeta_k -  \frac{\mathcal{H}}{\bar\chi^\prime} \delta\chi_k \;
\end{equation}
with the conformal Hubble expansion rate $\mathcal{H}= a H$. Since this definition does not contain any time derivatives of perturbation variables, this change of variable from the original variables (\cref{eq:def_R}) amounts to a trivial canonical transformation. After eliminating all the other (non-dynamical) variables, we find the quadratic action for $\mathcal{R}_k$ as
\begin{equation}
\label{eq:38}
    S_S^{(2)} = \frac{\Mpl^2}{2}\int \dd \tau\, \dd^3k  \, z^2 \left(  \vert \mathcal{R}_k^\prime \vert^{2} - c_\mathcal{R}^2 k^2 \vert \mathcal{R}_k\vert^2  \right)\;,
\end{equation}
where
\begin{align}
    z^2 =& a^2 \alpha (1+w) \, \frac{k^2 +  \frac{3}{2} (1+w) \alpha \mathcal{H}^2 }
  {  c_s^2 \left( k^2 + \frac{3}{2} (1+w) \alpha \mathcal{H}^2 \right) + \frac{1+w}{2} \alpha \mathcal{H}^2 \left( \frac{1+w}{2} \alpha - \epsilon \right)}\,, \\
  c_\mathcal{R}^2 =& \frac{c_s^4 (1+w)^2 k^4 + B_1 \mathcal{H}^2 k^2 + B_2 \mathcal{H}^4 }{c_s^2 (1+w)^2 k^4 + A_1 \mathcal{H}^2 k^2 + A_2 \mathcal{H}^4}\; \label{eqn:def-cR2}
\end{align}
with
\begin{align}
    A_1 =& \frac{1}{4} (1+w)^3 \alpha  (12 c_s^2 + (1+w) \alpha - 2 \epsilon)\;, \\
    A_2 =& \frac{3}{8} (1+w)^4 \alpha^2 (6 c_s^2 + (1+w) \alpha - 2 \epsilon)\;, \\
    B_1 =& \frac{1}{4} c_s^2 (1+w)^2 \left( (1+w)^2 \alpha^2  + 6 (1+w) \alpha  (1+ 3 c_s^2 - \epsilon) + 4 \epsilon \eta \right)\;, \\
    B_2 =& \frac{1}{8} (1+w)^3 \alpha \big[ - (1+w)^2 \alpha^2 + 2 (1+w) \alpha (6 c_s^2 +9 c_s^4 + (2-3 c_s^2) \epsilon) + 4 \epsilon (- (1+ 3 c_s^2) \epsilon + 3 c_s^2 (1+ 3 c_s^2 + \eta) ) \big]
\end{align}
and
\begin{equation}
    \alpha = \frac{\rho_\chi a^2}{\mathcal{H}^2}\;, \qquad
    \epsilon = 1- \frac{\mathcal{H}^\prime}{\mathcal{H}^2}\;, \qquad
    \eta = \frac{\epsilon^\prime}{\epsilon \mathcal{H}}\;.
\end{equation}
The equation of motion is then simply given by
\begin{equation}
\label{eq:scalar_EOM}
    v_k^{\prime\prime} + \left(c_\mathcal{R}^2 k^2 - \frac{z^{\prime\prime}}{z}\right) v_k=0\;,
\end{equation}
where we have introduced the Mukhanov-Sasaki-type variable $v_k = z \mathcal{R}_k$. We note that the structure of $z$ and $c_\mathcal{R}$ appears to include non-local terms. However, in the ultraviolet limit $k\rightarrow \infty$ and with finite $\mathcal{H}$ as well as in the regime of GR at the background level, \ie. $\alpha (1+w)=2 \epsilon$ and $\eta = - 3 (1+w) + 2 \epsilon$, for all $k$, we recover the usual equations of motion from GR.

\section{Bouncing scenario}
\label{sec:bouncing}

\subsection{Set-up}
In order to search for a viable parameter space in which the scalar power spectrum is almost scale invariant, we first note that in the regime where the modified gravity from the potential $V(\phi)$ dominates, \ie.~$\alpha \mathcal{H}^2 / k^2 \ll 1$, the form of $z^2$ and $c_\mathcal{R}$ can be simplified to
\begin{equation}
    z^2 \approx a^2 \alpha \frac{(1+w)}{c_s^2}\;, \qquad c_\mathcal{R}^2 \approx c_s^2\;.
\end{equation}
Therefore, in that regime we can solve \cref{eq:scalar_EOM} approximately. Supposing that the scale factor behaves as $a \propto (\tau^2)^{n/2}$ we obtain
\begin{equation}
\label{eq:approximated_EOM_scalar}
    \frac{\dd^2v_k}{\dd x^2} + \left(c_s^2 \kappa^2 - \frac{n (3w-1) (-2 + n (3w -1))}{4 x^2} \right) v_k \approx 0 \; ,
\end{equation}
where we have introduced $x=\tau/\tau_B$ and $\kappa=k \tau_B$ with $\tau_B > 0$ the bouncing time scale. Therefore, at that regime the independent solutions are as usual given by the Hankel functions, provided $w$ and $c_s^2$ are constant. Assuming that the regime $\alpha \ll 1$ holds up to horizon crossing for the cosmological microwave background (CMB) scales we can estimate that the spectral index is given via 
\begin{equation}
    n (3w-1) (-2 + n (3w -1)) = 15 - 8 n_s + n_s^2\;,
\end{equation}
which yields the following two solutions for $n$
\begin{equation}
    \label{eq:n_in_terms_of_omega}
    \frac{5-n_s}{3w -1}\quad \text{and}\quad \frac{n_s -3}{3w -1}\;.
\end{equation}
In order to have a valid bouncing solution we require that $n>0$. Therefore, the first solution is valid for $w >1/3$ and the second one for $w<1/3$.

However, an equation of state $w< 1/3$ can lead to issues since the anisotropies then grow faster than the energy density of the scalar field in the contracting phase. That is why we shall focus on the first case with $w \geq 1$. Note that this corresponds to $z \propto \tau^{(-3+n_s)/2}$. For $n_s<3$, which is the case for the primordial curvature perturbation of our universe, $z$ increases in time during the contraction phase and, therefore, $\mathcal{R}_k$ has a decreasing and constant mode in contrast to common bouncing scenarios.

After the bounce we want to recover the usual relations from GR. This can be achieved either by considering a non-constant equation of state or a transition of the scale factor. We consider the latter case so that
\begin{equation}
    \lim_{\tau\rightarrow \infty} a \propto \tau^{\frac{2}{3w +1}}\;.
\end{equation}
In order to match the background after the bounce with GR we have to further ensure that $\lim_{\tau\rightarrow\infty}3 \mathcal{H}^2/(\rho_\chi a^2)=1$, which fixes the normalization $\rho_0$ of $\rho_\chi$. Combining these both solutions including the bounce we consider the following ansatz 
\begin{equation}
    \label{eq:Scale_factor}
    a (\tau) = a_0 \left( \frac{\tau^2}{\tau_e^2} \right)^{\frac{n}{2}} \Theta(\tau_e-\tau) + a_1 \left[ 1+ \left( \frac{\tau}{\tau_B} \right)^2 \right]^{\frac{1}{3w+1}} \Theta(\tau-\tau_e)\;,
\end{equation}
where $a_1=a_0 (1+(\tau_e/\tau_B)^2)^{-1/(3w+1)}$ to ensure continuity and the step function is operationally defined as
\begin{equation}
    \Theta(x) = \lim_{m\rightarrow \infty} \frac{1}{1+ e^{-m x}}\;.
\end{equation}
In the above, the time $\tau_e$ locates the transition between the two different regimes, which we place before the bounce, \ie. $\tau_e < 0$. For numerical purposes we have to choose a finite $m$. The bigger $m$ the sharper the transition, but this may also lead to numerical issues, since the derivatives start to diverge. Later on, we will actually choose rather small values of $m$.

Furthermore, depending on $n$, finite $m$ may again bring other numerical issues around $\tau=0$, at the bounce. That is why it may be convenient to slightly detune the relation by introducing a small $\tau_a / \tau_e$ shift. Explicitly, the ansatz of \cref{eq:Scale_factor} is modified to
\begin{equation}
    \label{eq:Scale_factor_detuned}
    a(\tau) = a_0 \left[ \left( \frac{\tau}{\tau_e} \right)^2 + \left( \frac{\tau_a}{\tau_e} \right)^2 \right]^{\frac{n}{2}} \Theta(\tau_e-\tau) + a_1 \left[ 1 + \left( \frac{\tau}{\tau_B} \right)^2 \right]^{\frac{1}{3w +1}} \Theta(\tau-\tau_e)
\end{equation}
with $a_1 = a_0  (1+(\tau_a/\tau_e)^2)^{n/2}(1+ (\tau_e/\tau_B)^2)^{-1/(3w+1)}$, where $|\tau_a| \ll |\tau_e|$.
The role of $\tau_a$ is only to regulate the behavior of (the derivatives of) $a$ at the bounce, and we
shall later check that the choice of $\tau_a$ with $|\tau_a| \ll |\tau_e|$ does not impact the final result.

\subsection{Reconstruction of $V(\phi)$}
\label{subsec:reconstruction}

Using the background equation of motion (\cref{eq:10}), we can solve $\phi$ in terms of the conformal time as
\begin{equation}
\phi = \frac{3}{2} \int \dd \tau^\prime a (1+ w) \rho_\chi + \phi_0\;,
\end{equation}
where $\phi_0$ is an integration constant. If we are to consider the full period, this equation can be solved numerically. Before doing so, let us first have a look at the two different regimes separately. 

On one hand, deep in the contraction phase, where $-\tau \gg |\tau_e|$ the scale factor is well approximated by
\begin{equation}
  a(\tau) \approx a_0 \left( \frac{\tau}{\tau_e} \right)^n\; ,
\end{equation}
in which $\phi$ becomes a function of time as,
\begin{equation}
    \label{eq:phi(tau)_before_the_bounce}
    \phi(\tau ) \approx \frac{3}{2} \frac{ (1+w) \, a_0 \, \rho_0}{1- n\, (2 + 3 w) } \left( \frac{\tau}{\tau_e}\right)^{ - n\, (2+ 3w)} \tau + \phi_0\;.
\end{equation}
The scalar field $\phi$ asymptotically approaches $\phi_0$ for $\tau \to - \infty$ and then grows monotonically in the contraction phase before the transition period. Using \cref{eq:const_1}, the potential can then be reconstructed as 
\begin{equation}
    \label{eq:Reconstructed_potential_before_the_bounce}
    V(\tau) =  \frac{1}{3} \left( \frac{3}{2} \frac{ (1+w) \, a_0 \, \rho_0}{1- n\, (2 + 3 w) } \left( \frac{\tau}{\tau_e}\right)^{ - n\, (2+ 3w)} \tau + \phi_0  \right)^2  - \rho_0 \left( \frac{\tau}{\tau_e} \right)^{-3 n \, (w+1)}\;.
\end{equation}
Since during this phase $\phi$ is monotonically increasing in time we can invert the relation \eqref{eq:phi(tau)_before_the_bounce} to express $\tau$ in terms of $\phi$ and $V$ in terms of $\phi$.

On the other hand, after the transition, but before the bounce for large $m\gg 1$  the scale factor behaves as
\begin{equation}
    a(\tau) \approx a_1 \left( 1 + \frac{\tau^2}{\tau_B^2} \right)^{\frac{1}{3w+1}}\;,
\end{equation}
which leads to
\begin{equation}
    \phi(x) \approx \phi_0  + \frac{3}{2} \left( \frac{a_0}{a_1} \right)^{3(1+w)} (1+w) \, a_1 \tau_B \rho_0 x \,_{2}F_1\left( \frac{1}{2}, \frac{2+3w}{1+3w},\frac{3}{2}, -x^2 \right)\;,
\end{equation}
where $_{2}F_1$ denotes the hypergeometric function.
Similarly as before, the potential then reads
\begin{equation}
    V(x) \approx \frac{1}{3} \left[ \phi_0  + \frac{3}{2} \left( \frac{a_0}{a_1} \right)^{3(1+w)} (1+w) \, a_1 \tau_B \rho_0 x \,_{2}F_1\left( \frac{1}{2}, \frac{2+3w}{1+3w},\frac{3}{2}, -x^2 \right) \right]^2 - \rho_0 \left( \frac{a_0}{a_1} \right)^{3(1+w)} \left( 1 + x \right)^{- \frac{3(1+w)}{1+3w}} \;.
\end{equation}

\begin{figure}
    \centering
    \includegraphics[width=8.255cm]{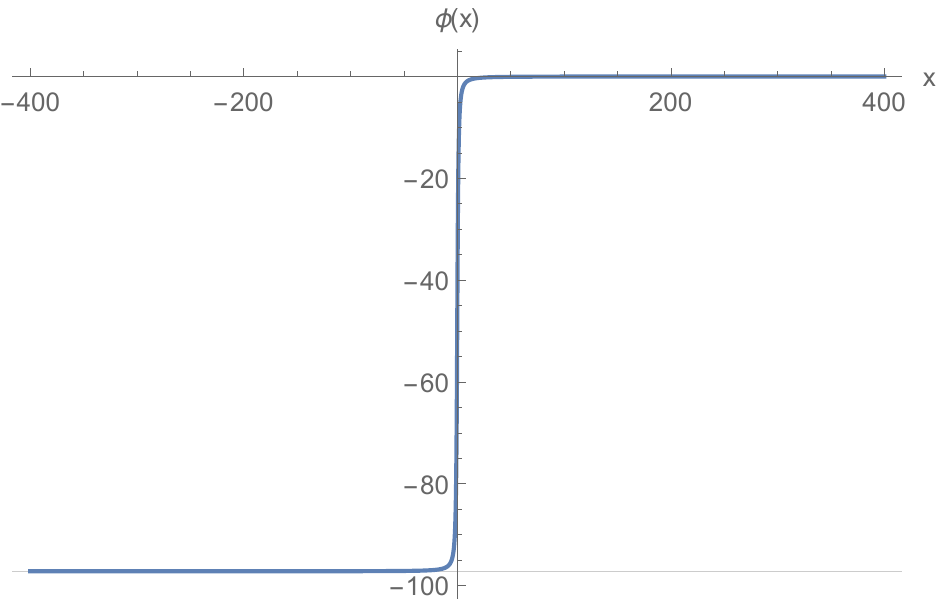}\hfill
    \includegraphics[width=8.255cm]{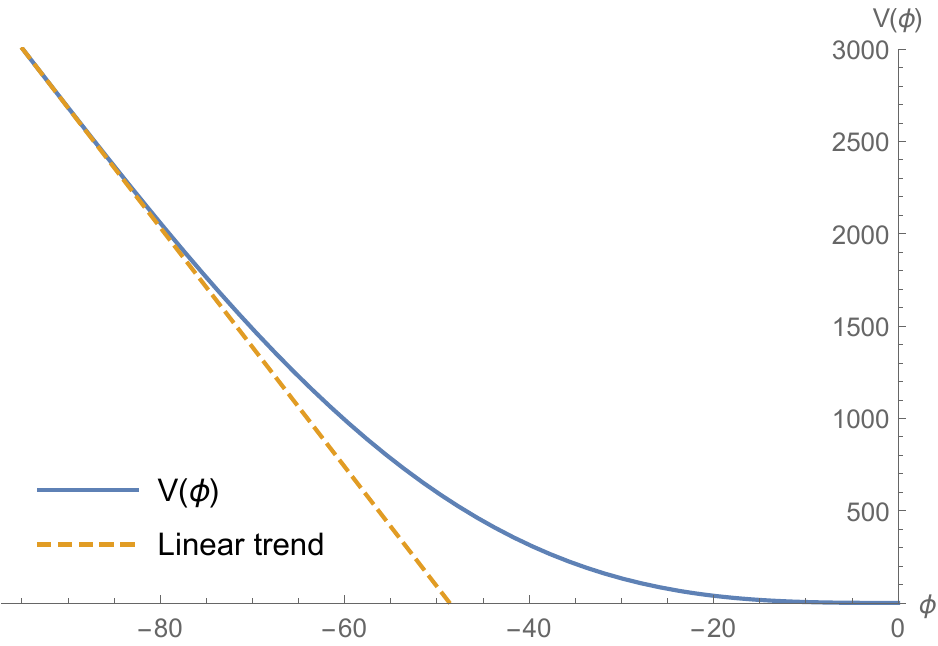}
    \caption{The form of the reconstructed potential for $w=1$, $n=2.02$, $m=1$, $\tau_e=-300\tau_B$ and $\tau_a=0$ (right) and the evolution of the scalar field $\phi(x)$ for the same parameter set (left), where $x=\tau/\tau_B$. Notice that the linear trend indicated (right) is here simply built from the tangent at the minimum of $\phi$.}
    \label{fig:Reconstructed_Potential}
    \label{fig:Evolution_Phi}
\end{figure}

In \cref{fig:Evolution_Phi}, we give the result obtained by the numerical simulation of $\phi(x)$ and $V(\phi)$, across the bounce. The left-hand side plot shows the evolution of the scalar field $\phi(x)$ for the case where $w=1$, $n=2.02$, $m=1$, $\tau_e=-300\tau_B$ and $\tau_a=0$, which gives $n_s = 0.96$. We there observe that the scalar field is indeed monotonically growing. Therefore, we can invert $\phi(x) \rightarrow x(\phi)$ to reconstruct the potential $V(\phi)$ which is given in the right-hand side plot of \cref{fig:Reconstructed_Potential}. We choose the integration constant $\phi_0$ such that $\phi$ goes to $0$ in the limit $x \to \infty$, and then we see $V \to 0$ in the same limit. Before the bounce but after the transition, \ie. $-300 (=\tau_e/\tau_B) \ll x \ll -1$, the potential approaches a linear trend. This is expected since for $V(\phi) \propto \phi$, we recover GR. For $x \ll -300 (=\tau_e/\tau_B)$ the potential (\cref{eq:Reconstructed_potential_before_the_bounce}) models the impact of matter and is therefore expected to deviate from the linear trend. However, since the scalar field is roughly constant in that regime the deviation is not visible anymore on the plot. 

\subsection{Power spectrum}
\label{subsec:power}

\subsubsection{Scalar part}
\label{subsubsec:scalar_power}

We first consider the scalar part and solve the equation of motion \eqref{eq:scalar_EOM} numerically. In the present study, we fix the transition time scale $\tau_e=-300 \tau_B$ and choose a slow transition with rather small values of $m$ for computational ease. Note that in our convention the bouncing time scale $\tau_B$ is taken to be strictly positive.

\begin{figure}[ht]
    \includegraphics[scale=0.65]{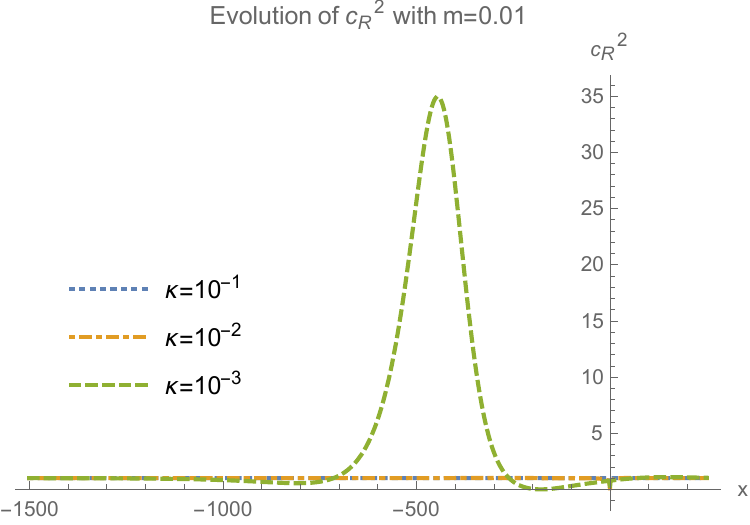} \hfill
    \includegraphics[scale=0.65]{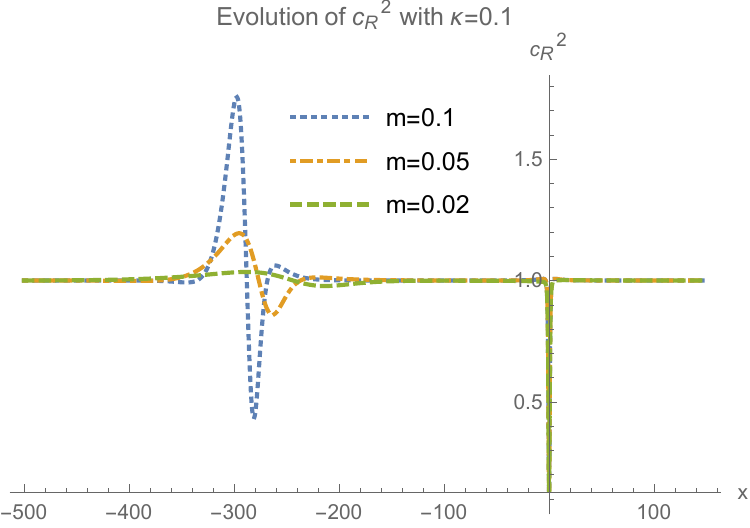} \\
    \vspace{5pt}

    \includegraphics[scale=0.65]{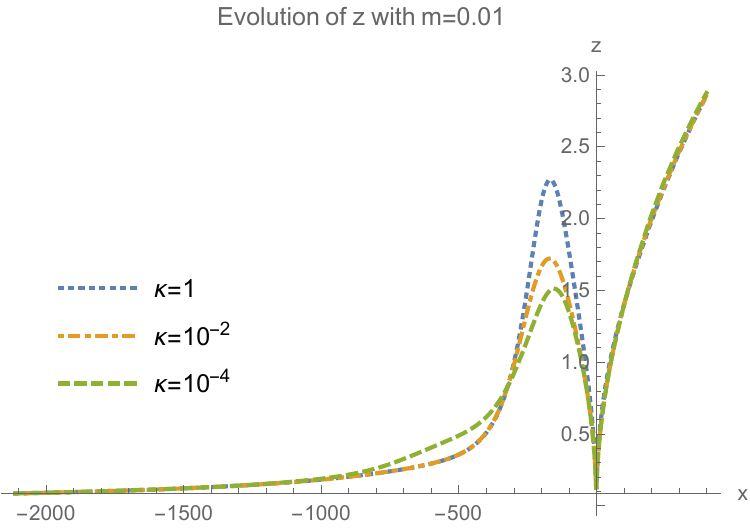} \hfill
    \includegraphics[scale=0.65]{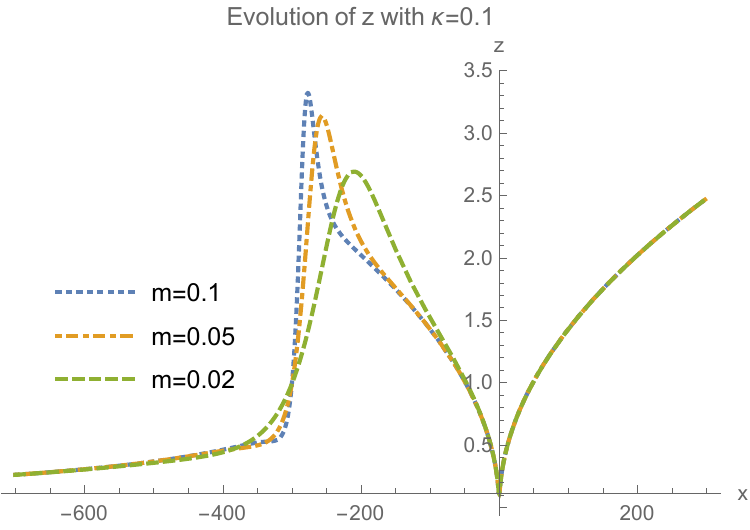}
    \caption{In the upper panels we plot the sound speed square for different values of $\kappa$ (top left) and $m$ (top right). In the lower panels, we plot $z$ for, again, different values of $\kappa$ (bottom left) and $m$ (bottom right). For these plots we consider the case of $n=2.02$ and $w=1$.}
    \label{fig:sound_speed_and_z}
\end{figure}

In \cref{fig:sound_speed_and_z}, we plot the sound speed square $c_\mathcal{R}^2$ and $z$, for $n=2.02$ and $w=1$ (following \cref{eq:n_in_terms_of_omega}) for different values of $\kappa = k\tau_B$ and $m$.
We can observe that $z$ remains positive throughout the evolution and, as expected, it remains independent of $m$ and $\kappa$ both at very early times and after the bounce. However, in the regime around the transition, at $x=-300$, $z$ depends both on $m$ and $\kappa$. In particular for very small values of $\kappa$, the value of $z$ starts to deviate earlier from the approximated behavior $z^2 \approx a^2 \alpha (1+w)/c_s^2 $. This can be easily understood: the approximation is only valid for a large ratio of $k^2=\kappa^2/\tau_B^2$ to $\alpha \mathcal{H}^2$.

The behavior of $c_\mathcal{R}^2$, which is defined in \cref{eq:38,eqn:def-cR2}, is similar. It deviates only around the transition regime, \ie.~when the dependency of $m$ actually manifests itself. Again, the dependency on $\kappa$ depends on the ratio $k^2 / (\alpha \mathcal{H}^2)$. Note that for $\kappa \ll 1$ the sound speed square $c_\mathcal{R}^2$ can become negative around the transition regime. 
However, this does not correspond to the standard gradient instability with the exponentially fast growth in the ultraviolet (UV), since in the UV limit ($\kappa \gg 1$) the sound speed squared given in \cref{eqn:def-cR2} is positive-definite. On the other hand, in the infrared (IR) or at large scales ($\kappa \ll 1$), where $c_\mathcal{R}^2 <0$, the frequency is still positive-definite since $\vert z^{\prime\prime}/z \vert \gg \vert c_\mathcal{R}^2 k^2 \vert$ (and $z^{\prime\prime}/z < 0$) so that $\omega^2 \equiv c_\mathcal{R}^2 k^2 - z^{\prime\prime}/z >0$. Therefore, the model is not plagued by either UV or IR instabilities during the transition phase. Furthermore, the model should be free from the strong coupling, which is usually~\footnote{This is indeed the case \eg. in the framework of EFT of single-field inflation/dark energy, see \eg.~\cite{Cheung:2007st}.} signaled by vanishing of the UV/subhorizon (\ie. $\kappa\gg 1$) sound speed and which is insensitive to the dispersion relation in the intermediate/IR scales. Each mode remains within the regime of validity of the perturbative expansion and smoothly evolves from the initial time to the final time.

\begin{figure}[ht]
    \centering
    \includegraphics[scale=0.65]{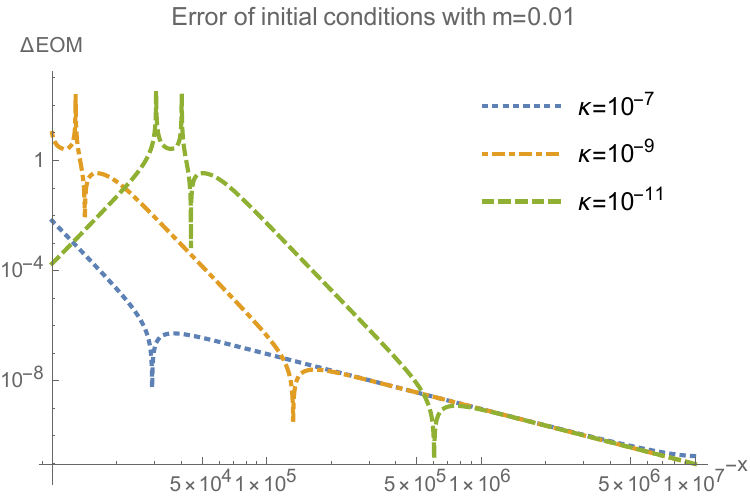} \hfill
    \includegraphics[scale=0.65]{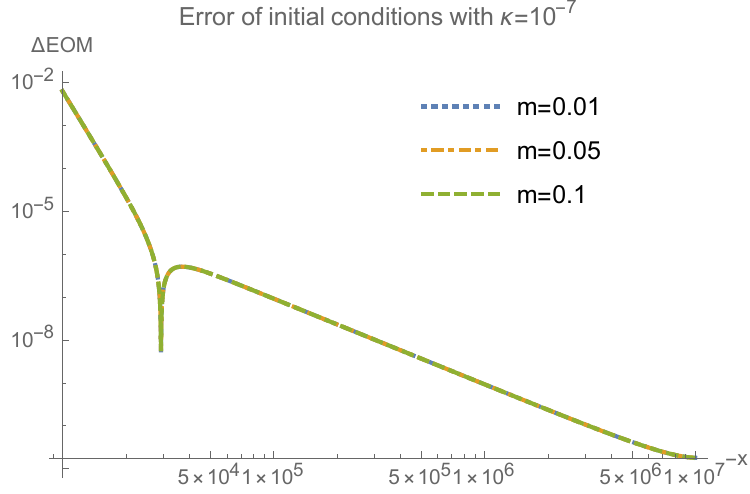}
    \caption{The normalized error of the analytical solution for $n=2.02$ and $w=1$, for different values of $\kappa$ with $m=0.01$ (left) and for different values of $m$ with $\kappa=10^{-7}$ (right).}
    \label{fig:Error_EOM}
\end{figure}
In order to numerically obtain the scalar power spectrum, we fix the initial conditions to the standard adiabatic vacuum so that
\begin{equation}
   \frac{ v_k (x=x_i)}{\sqrt{\tau_B}} = \frac{\sqrt{\pi}}{2} \sqrt{-x_i} H_{\frac{4-n_s}{2}}^{(1)}\left(-c_s \kappa x_i\right)
   \label{eq:analytical_Solution}
\end{equation}
and similarly for its derivative. Firstly, we check that our initial conditions for $x \ll -300$ are indeed valid. In \cref{fig:Error_EOM} we plot the normalized error $\Delta \mathrm{EOM}$ of the initial conditions for different values of $\kappa$ or $m$, that is the quantity 
\begin{equation}
    \Delta \mathrm{EOM} = \Big \vert \frac{1}{v_k}\frac{\dd^2 v_k(x)}{\dd x^2}  \left( c_\mathcal{R}^2 \kappa^2 - \frac{1}{z}\frac{\dd^2 z}{\dd x^2} \right)^{-1} + 1 \Big \vert\;.
\end{equation}
We verify that far away from the transition regime where $\alpha \mathcal{H}^2 / k^2 \ll 1$ the error is negligibly small. It only starts to increase during the contraction phase, just as expected. For smaller values of $\kappa$, the impact of the scale-dependent mass and sound speed starts to matter earlier since the approximation depends on the ratio of $\alpha \mathcal{H}^2$ to $k^2=\kappa^2/\tau_B^2$. On the other hand, changing the value of $m$ does not have any impact at early times. Thanks to the good agreement at $x \leq - 10^6$, we do not need to start evolving the EOM from inside the horizon. We can instead start outside the horizon using the analytic approximation. In the following we will fix the starting point for our numerical solution at $x=-5\cdot 10^6$ for $10^{-11}\leq \kappa \leq 10^{-7}$. Different starting values do not affect the conclusions of this work.
From thereon, we shall use $m=0.01$.

\begin{figure}
    \includegraphics[scale=0.65]{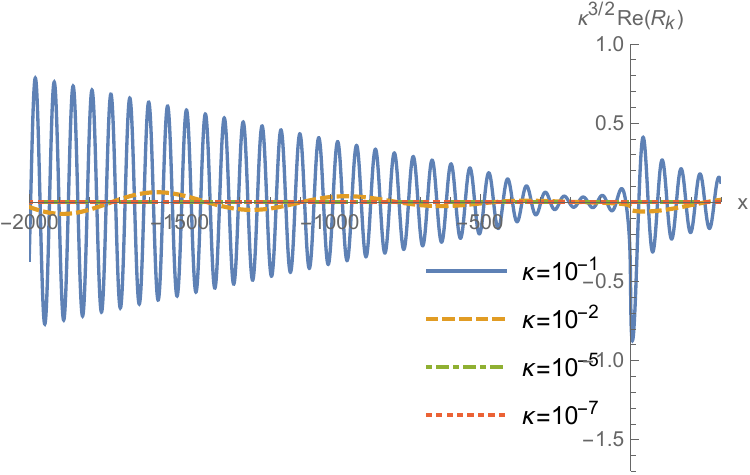} \hfill
    \includegraphics[scale=0.65]{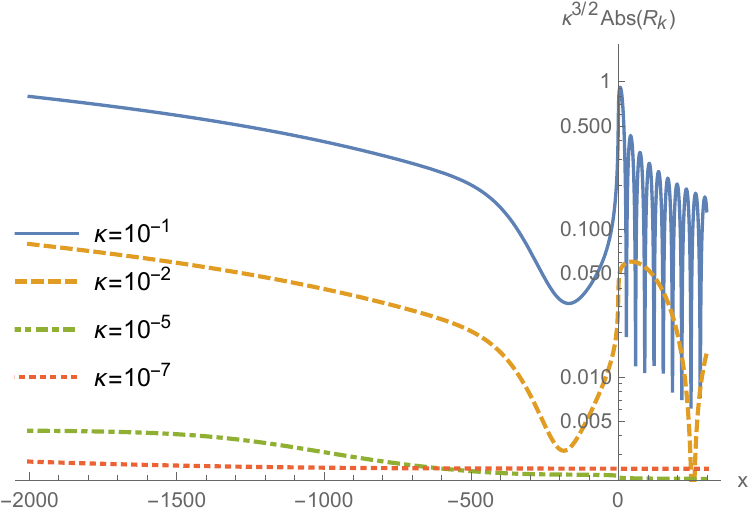}
    \caption{The real part (left) and the absolute value (right) of the normalized evolution of the curvature perturbation modes for $n=2.02$ and $w=1$, for different values of $\kappa$.}
    \label{fig:Evolution_Curvature modes}
\end{figure}

\begin{figure}
    \includegraphics[scale=0.65]{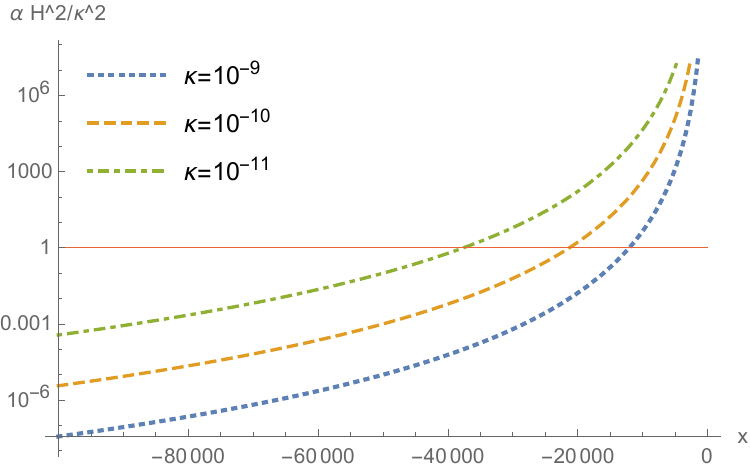}
    \includegraphics[scale=0.65]{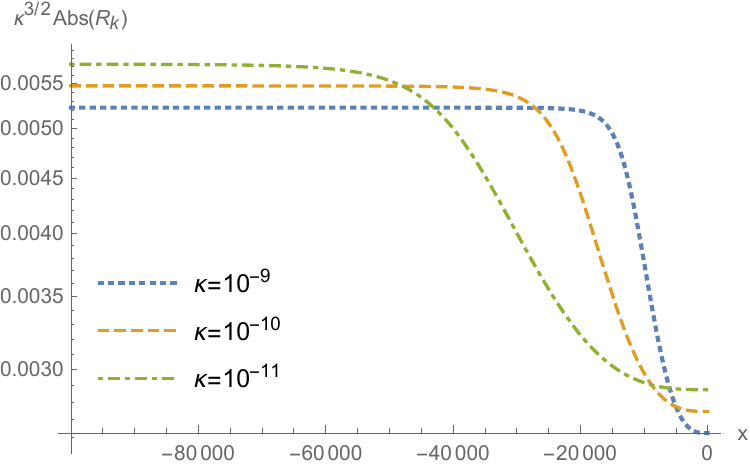}
    \caption{The left-hand side displays the ratio $\alpha \mathcal{H}^2/k^2$, and the right-hand side shows the normalized absolute value of the curvature perturbation for $n=2.02$ and $w=1$.}
    \label{fig:alpha_Curvature}
\end{figure}

In \cref{fig:Evolution_Curvature modes}, we exhibit the real part and the absolute value of $\kappa^{3/2} \mathcal{R}_k$ for different values of $\kappa$.
We see that neither at the transition regime nor at the bounce, do we obtain any instability.
In fact, neither the transition nor the bounce has any significant impact on the curvature perturbation modes which are already far outside the horizon. For large $\kappa$ the comoving curvature perturbation is oscillating, while for small $\kappa$ the curvature perturbation is frozen. However, we still have to be careful. Our analytic approximation holds as long as $\alpha \mathcal{H}^2 \ll k^2$ ($=\kappa^2/\tau_B^2$). In \cref{fig:alpha_Curvature}, the left-hand side plot shows the ratio for small values of $\kappa$. For small values of $\kappa$, the approximation here breaks down outside the horizon but still far away from the bouncing regime. The right-hand side plot shows the normalized absolute value of the curvature perturbation (similarly to the right-hand side of \cref{fig:Evolution_Curvature modes}). As expected, the curvature perturbation is frozen before the breakdown of our analytic approximation. During the regime where $\alpha \mathcal{H}^2/ k^2 \approx 1$ the curvature perturbation falls down before freezing again. Therefore, the curvature power spectrum after leaving the horizon does not coincide with the one after the bounce. 

\begin{figure}[ht]
    \includegraphics[scale=0.65]{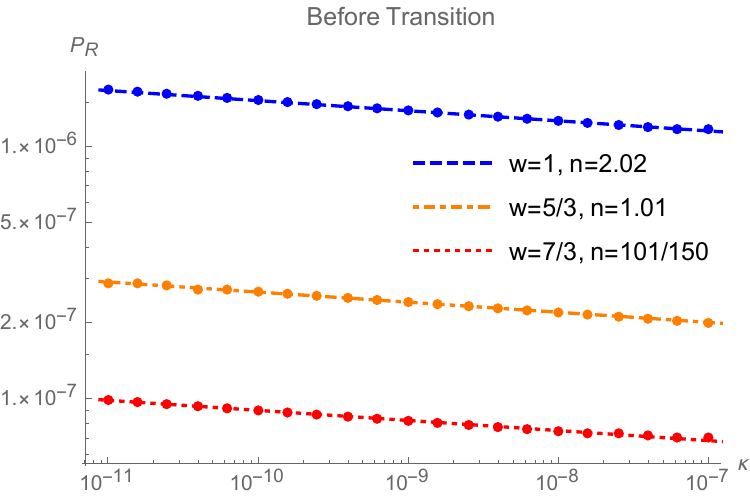} \hfill
    \includegraphics[scale=0.65]{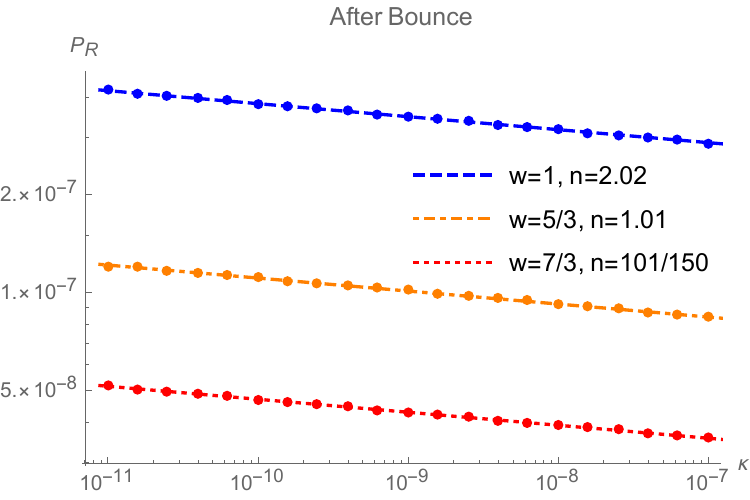}
    \caption{The scalar power spectrum before (left) the transition period and after the bounce (right) for different combinations of $w$ and $n$, \ie. $w=1$ \& $n=2.02$, $w=5/3$ \& $n=1.01$ and $w=7/3$ \& $n=101/150$.  }
    \label{fig:Scalar_Power_spectrum}
\end{figure}

In \cref{fig:Scalar_Power_spectrum}, the power spectrum is given before (at $x=-10^6$) and after the bounce (at $x=300$) for different combinations of $w$ and $n$, along with their fit by a spectral index of the form $A \cdot \kappa^{n_s-1}$, where $A$ is the amplitude of $P_{\mathcal{R}} \equiv k^3 \vert \mathcal{R}_k \vert^2/ (2\pi^2)$.
The power spectrum $P_{\mathcal{R}}$ is indeed slightly red-tilted with the correct spectral index of $n_s=0.96$ for the three different combination of $w$ and $n$, for small values of $\kappa$. The overall amplitude is slightly different for the curvature power spectrum, while the spectral shape is the same before and after the bounce, because of the aforementioned fall and freezing behavior.

\subsubsection{Tensor part}
\label{subsubsec:tensor_power}

The tensor perturbations are the same as in GR and are governed by the equation
\begin{equation}
    \frac{\dd^2 u_k}{\dd x^2} + \left( \kappa^2 - \frac{1}{a}\frac{\dd^2 a(x)}{\dd x^2}   \right) u_k=0\;.
\end{equation}
Therefore, it only depends on the specific form of the scale factor. Before the transition period the scale factor is given by $a \propto (-x)^n$. Therefore, for $n\neq 1$, the solutions are given by the Hankel functions
\begin{equation}
  \frac{u_k}{\sqrt{\tau_B}} = \frac{\sqrt{\pi}}{2}\sqrt{-x} H_\nu^{(1)}(-\kappa x)\;, 
  \quad\text{where}\quad
  \nu = \frac{2n-1}{2} \;.
\end{equation}
The tensor power spectrum for the cosmological scales will explicitly depend on $n$ and is not anymore always almost scale invariant, but instead we have
\begin{equation}
    \label{eq:nt_wrt_n}
    n_t = 4-2n
\end{equation}
for $n>1/2$, at horizon crossing in the contracting phase. Here $n_t$ is the spectral index of the tensor power spectrum $P_h \equiv \sum_\sigma k^3 \vert h_\sigma \vert^2 / (2\pi^2) = A_h k^{n_t}$, with $A_h$ being its amplitude. Therefore, among the three different cases considered for the scalar part with $n=2.02$, $n=1.01$ and $n=101/150 \approx 0.67333\cdots$, only the first case leads to an almost scale invariant power spectrum. The other ones are blue tilted.

\begin{figure}
    \centering
    \includegraphics[scale=0.65]{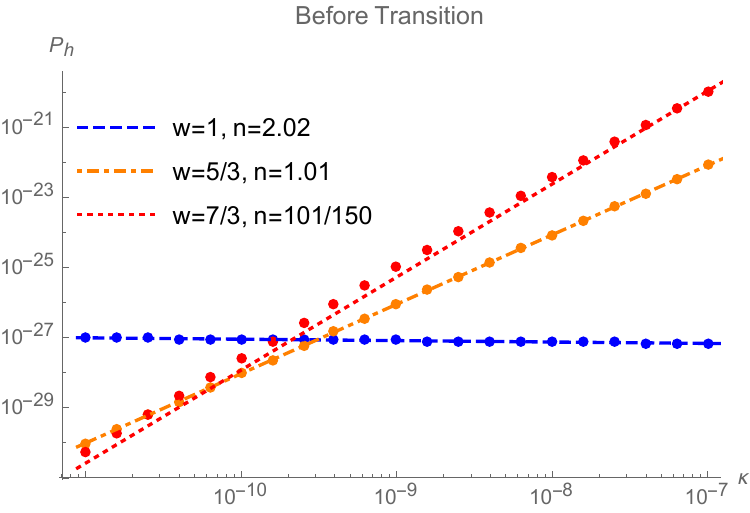} \hfill
    \includegraphics[scale=0.65]{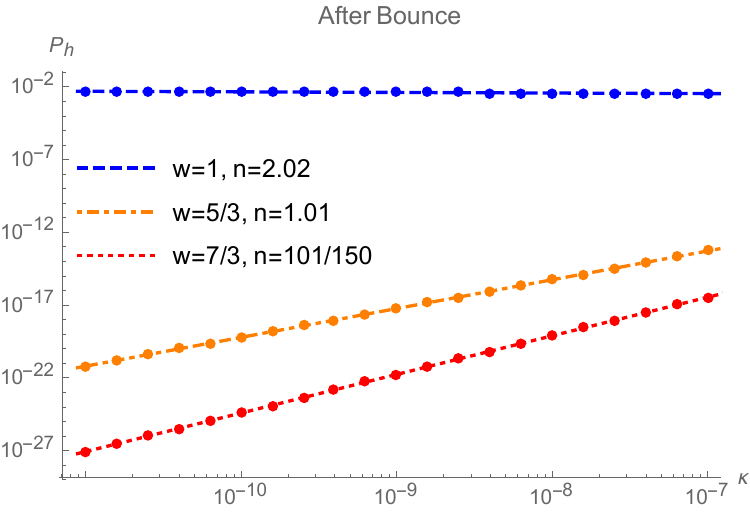}
    \caption{Tensor power spectra evaluated at $x=-3000$ before the transition (left) and at $x=300$ after the bounce (right) are plotted for different combinations of $w$ and $n$, \ie. $w=1$ \& $n=2.02$, $w=5/3$ \& $n=1.01$ and $w=7/3$ \& $n=101/150$.}
    \label{fig:Tensor_power_spectrum}
\end{figure}

The scale factor is decreasing in the contracting phase and, therefore, outside the horizon the tensor modes are either frozen or growing, in contrast to the scalar modes.
\begin{figure}
    \includegraphics[scale=0.65]{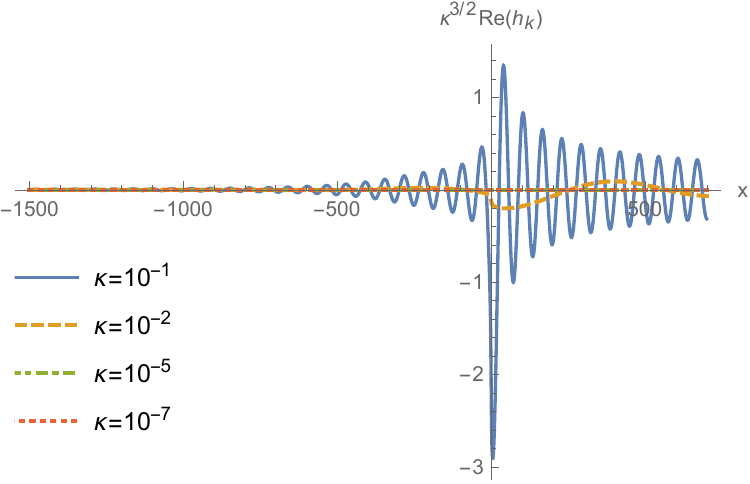} \hfill
    \includegraphics[scale=0.65]{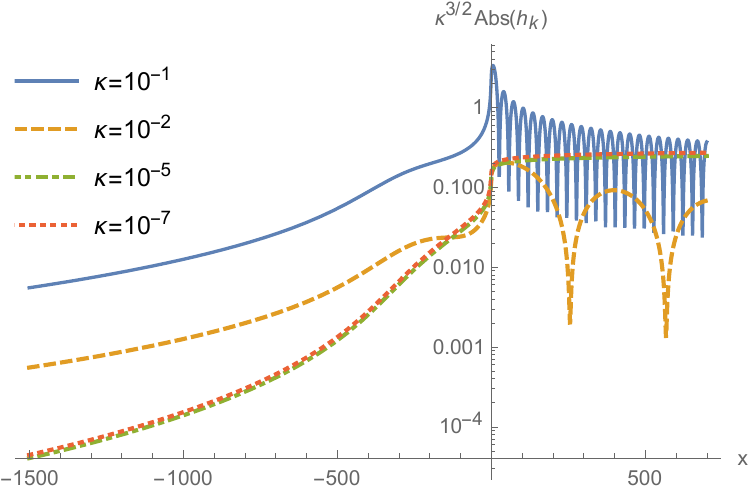}
    \caption{The real part (left) and the absolute value (right) of the tensor modes are plotted for $n=2.02$, $w=1$ and different values of $\kappa$.}
    \label{fig:Evolution_tensor_modes}
\end{figure}
The time evolutions of the tensor modes for $n=2.02$ are given in \cref{fig:Evolution_tensor_modes} and we observe that on superhorizon scales the modes are indeed growing. However, as for the scalar modes, neither the bounce nor the transition period impacts the scale dependency of the tensor power spectrum, as \cref{fig:Tensor_power_spectrum} demonstrates. Instead, it only leads to an overall amplification factor coming from the superhorizon growth in the contracting phase. 
Only for $n=2.02$ do we recover the almost scale invariant power spectrum. In fact, the spectral index of the tensor and scalar modes are the same in this case, which renders the comparison straightforward. 

However, in that case, the tensor-to-scalar ratio, \ie. $r \equiv P_h / P_\mathcal{R}$, is extremely large ($r \gg 1$), making this option unviable. This is apparent when comparing \cref{fig:Scalar_Power_spectrum,fig:Tensor_power_spectrum}. On the other hand, for $n<2$ the tensor spectrum is blue tilted and the tensor-to-scalar ratio becomes scale dependent. Indeed, one can write the tensor-to-scalar ratio as 
\begin{equation}
    r = r_0 \kappa^{n_t-n_s+1} \;.
\end{equation}
Assuming $n_s \approx 0.96$ and using \cref{eq:nt_wrt_n}, the same conclusion is easily drawn. Numerically, it translates as $r \propto \kappa^{1.94}$ for $n=1.01$ or $r \propto \kappa^{2.61333\cdots}$ for $n=101/150$. In these two cases, on cosmological scales, the tensor power spectrum is significantly lower than the scalar one $r\vert_{k=k_{CMB}} \ll 1$ as long as the time scale of the bounce is significantly shorter than the scale at the CMB, \ie. $\kappa_{\text{CMB}}= k_{\text{CMB}} \tau_B \ll 1$. Practically, this latter assumption should be easily satisfied.

\section{Discussion and conclusion}
\label{sec:conclusion}

In this study, we introduced an explicit and testable bouncing universe scenario, built within the framework of minimally modified gravity theories, specifically the class of so-called VCDM models. The proposed model successfully passes the first tests a bounce scenario has to face. It does not suffer from ghost or gradient instabilities coming from the null-energy condition violation and there are no issues related to the anisotropic stress or the BKL instability thanks to the ekpyrotic ($w \geq 1$) equation of state. From the observational side, the scalar power spectrum can be adapted by the choice of the equation of state and the form of the potential leading to a nearly scale-invariant power spectrum with a spectral index of $n_s \approx 0.96$ in accordance of the results of the Planck collaboration \cite{Planck:2018vyg}. Moreover, the tensor power spectrum scales, in general, differently from the scalar one. An equation of state $w >1$ leads to a blue tensor spectrum. It is, therefore, possible to obtain a small tensor-to-scalar ratio within the observational bounds at cosmological scales, while potentially detectable at much smaller scales such as those of the gravitational-wave interferometers. To meet all these goals, the current model relies on a simple asymmetric bounce with the minimal number of propagating dof's ($1$ scalar $+$ $2$ tensors), unlike previous works based on Cuscuton \cite{Boruah:2018pvq,Kim:2020iwq}, in which case the authors introduced an additional scalar field to fulfill the experimental constraints. This is a key part of this work: we have built our model based on the VCDM, which can accommodate both modified gravity behavior and GR behavior, and have reconstructed the potential in the Lagrangian from the background dynamics we chose.

Future work could investigate how sensitive to the bounce details (\eg. shape, duration, etc...) these tests are. \textit{A priori}, we argue that the conclusion of this work should prove relatively robust in this regard. Another crucial aspect to consider would be non-Gaussianities, and evading the no-go theorem associated with it \cite{Akama:2019qeh,Li:2016xjb}. However, since the curvature perturbations are frozen outside the horizon, non-Gaussianities are generated inside the horizon when the kinetic energy of the scalar field is subdominant. Naively, this should lead to small non-Gaussianities \cite{Bartolo:2021wpt}. Otherwise, one may also worry of seeing a superluminal sound speed in the matter sector ($k$-essence field). A standard ekpyrotic scalar field may sooth this, but would require a more complicated, and probably more numerically-involved, approach to handle.

\acknowledgments
A.G. receives support by the grant No. UMO-2021/40/C/ST9/00015 from the National Science Centre, Poland. P.M. acknowledges support from the Japanese Government (MEXT) scholarship for Research Student. The work of S.M. was supported in part by Japan Society for the Promotion of Science Grants-in-Aid for Scientific Research No. 17H02890, No. 17H06359, and by World Premier International Research Center Initiative, MEXT, Japan.
R.N.~was in part supported by the RIKEN Incentive Research Project grant.
A.G. wishes to thank YITP for hospitality during the development of this project. 

\bibliography{bibliography}

\end{document}